\documentclass{svmult}
\usepackage{graphicx}
\begin{document}
\title*{Modified gravity without dark matter}
\author{R.H. Sanders}
\institute{Kapteyn Astronomical Institute, Groningen, The Netherlands}
\maketitle
\def\ho{$H_o\,\,$}
\def\apj{{\it Astrophys.J.} }
\def\apjl{{\it Astrophys.J.Lett}}
\def\aj{{\it Astron.J.}} 
\def\mnras{{\it Mon.Not.RAS}}
\def\aa{{\it Astron.Astrophys.}}
\def\pr{{\it Phys.Rev.}}
\begin{abstract}
On an empirical level, the most successful alternative to
dark matter in bound gravitational systems is the modified Newtonian
dynamics, or MOND, proposed by Milgrom.
Here I discuss the attempts to formulate 
MOND as a modification of General Relativity.
I begin with a summary of the phenomenological successes of MOND and
then discuss the various covariant theories that have been proposed as
a basis for the
idea.  I show why these proposals have led inevitably to a 
multi-field theory.  I describe in some detail TeVeS, the 
tensor-vector-scalar theory proposed by Bekenstein, and discuss its 
successes and shortcomings.  This lecture
is primarily pedagogical and directed to those with some, but not a deep,
background in General Relativity 
\end{abstract}

\section{Introduction}

There is now compelling observational support for a standard 
cosmological model.  It is most impressive
that this evidence is derived from very different 
observational techniques applied to very different phenomena:  
from precise measurements of anisotropies in the
Cosmic Microwave Background (CMB) \cite{speal03}; from systematic
photometric observations of the light curves of distant supernovae
\cite{pereal99,garn,tonry}; from redshift surveys mapping
the distribution of observable matter on large scale and interpreting
that distribution in the context of structure formation by 
gravitational collapse \cite{sanch,eisen}.  Using the standard parameterised 
Friedmann-Robertson-Walker models (FRW), all of these observations imply 
a convergence
to a narrow range of parameters that characterise the Universe; this
convergence is rightly heralded as a remarkable achievement of the past decade.

However, the Universe that we
are presented with is strange in its composition:  only five percent
is the ordinary baryonic matter that we are familiar with; twenty-five
percent consists of pressureless dark matter presumed to
be fundamental particles that are as yet undetected by other means; and
about seventy percent is the even stranger negative pressure dark energy,
possibly identified with a cosmological term in Einstein's field equation,
and emerging relatively recently in cosmic history as the
dominate contributer to the energy density budget of the Universe.

A general sense of unease, primarily with this dark energy, has led
a number of people to consider the possibility that gravity may
not be described by standard four-dimensional General Relativity (GR)
on large scale (see Sami, this volume)-- that is to say, perhaps
the left-hand-side rather than the right-hand-side of the Einstein equation 
should be reconsidered.  Various possibilities have been proposed--
possibilities ranging from the addition of a 
scalar field with a non-standard kinetic term, K-essence \cite{apms};
to gravitational actions consisting of general functions of the usual
gravitational invariant, $F(R)$ theories (Sotiriou, this volume and 
\cite{cct03,ceal}); 
to braneworld scenarios
with leakage of gravitons into a higher dimensional bulk (\cite{dvali} and
Maartens, this 
volume).  But, in fact, there is a longer history of modifying
gravity in connection with the dark matter problem-- primarily
that aspect of the problem broadly described as ``missing mass'' 
in bound gravitational systems such
as galaxies or clusters of galaxies.  The observations of this
phenomenology have an even longer history, going back to the discovery
of a substantial discrepancy between the dynamical mass and the
luminous mass in clusters of galaxies \cite{zwk33}.
The precise measurement of rotation curves of spiral galaxies in
the 1970's and 1980's, primarily by 21 cm line observations which extend 
well beyond the visible disk of the galaxy \cite{Bos78,bg89}, demonstrated 
dramatically that this discrepancy is also present in galaxy systems.  

A fundamental, often implicit, aspect of the cosmological paradigm is that 
this observed discrepancy
in bound systems is due to the cosmological dark matter--
that the cosmological dark matter clusters on small scale and
promotes the formation of virialized systems via gravitational collapse
in the expanding Universe.  The necessity
of clustering on the scale of small galaxies implies that there are no
phase space constraints on the density of the dark matter and, hence, that
it is cold, or non-relativistic at the epoch of matter-radiation 
equality \cite{bdsl83}.  
The exact nature of the hypothetical cold dark matter 
(CDM) is
unknown but particle physics theory beyond the standard model provides
a number of candidates.  There are observational problems
connected with the absence of phase space constraints in this
dark matter fluid, problems such as the formation
of numerous but unseen satellites of larger galaxies \cite{deluc} 
and the prediction
of cusps in the central density distributions of galaxies-- cusps
which are not evident in the rotation curves \cite{debeal}.  But it is usually
taken as a article of faith that ``complicated astrophysical processes''
such as star formation and resulting feed-back will solve these problems.

The motivation behind considering modifications of
gravity as an alternative to CDM is basically the same as that underlying
modified gravity as an alternative to dark energy:  when a theory,
in this case GR, requires the existence of a medium which has not been,
or cannot be, 
detected by means other than its global gravitational influence, i.e., 
an ether, then it is not unreasonable to
question that theory. The primary driver for such
proposals has been the direct
observation of discrepancies in bound systems-- galaxies and clusters
of galaxies-- rather than cosmological considerations, such as that of 
structure formation in an expanding Universe.  The most successful
of the several suggestions,
modified Newtonian dynamics or MOND, has  
an entirely phenomenological rather than theoretical basis \cite{m83a,
m83b,m83c}.
In accounting for the detailed kinematics of galaxies and galaxy groups, 
while encompassing global scaling relations and empirical photometric
rules, MOND has, with one simple formula and one new fixed parameter, 
subsumed a wide range of apparently disconnected phenomena. 

In this respect it is similar to the early
proposal of continental drift by Alfred Wegener in 1912.  This 
suggestion explained a number of apparently disconnected geological
and palaeontological facts but had no basis in deeper theory; no one,
including Wegener, could conceive of a mechanism by which giant land
masses could drift through the oceans of the earth.  Hence the idea
was met with considerable ridicule by the then contemporary community of 
geologists and relegated to derisive asides in introductory textbooks. 
It was decades later, after the development of the modern theory
of plate tectonics and direct experimental support provided by 
the frozen-in magnetic field reversals near mid-oceanic
rifts, that the theory underlying continental drift 
became the central paradigm of geology and recognised as the principal
process that structures the surface of the earth \cite{wil}.
I do not wish to draw a close analogy between MOND and the
historical theory of continental drift, but only to emphasise the
precedent:  an idea can be basically correct but not generally accepted
until there is an understandable underlying physical mechanism-- until
the idea makes contact with more familiar physical concepts.

The search for a physical mechanism underlying modified Newtonian
dynamics is the subject here.  I begin with a summary of
the phenomenological successes of the idea, but, because this has
been reviewed extensively before \cite{sm02}, 
I will be brief.  I consider
the proposals that have been made for modifications of 
GR as a basis of MOND.  These proposals have led to the current best 
candidate-- the 
tensor-vector-scalar (TeVeS) theory of Bekenstein \cite{jdb04}, a theory that
is complicated but free of obvious pathologies. I summarise the successes
and shortcomings of the theory, and I present an
alternative form of TeVeS which may provide a more natural basis to the 
theory.  I end by a discussion of more speculative possibilities. 

\section{The phenomenology of MOND}
\subsection{The basics of MOND}
If one wishes to modify Newtonian gravity in an ad hoc manner in order
to reproduce an observed property of galaxies, 
such as asymptotically flat rotation curves, 
then it would seem most obvious to 
consider a $1/r$ attraction beyond a fixed length scale $r_0$.
Milgrom \cite{m83a} realized early on that this would not
work-- that any modification explaining the systematics of the
discrepancy in galaxies cannot be attached to a length scale but
to a fixed acceleration scale, $a_0$.  
His suggestion, viewed as a modification
of gravity, was that the true gravitational
acceleration ${\bf g}$ is related to the Newtonian gravitational acceleration
${\bf g_n}$ as
$$ {\bf g}\mu(|g|/a_o) = {\bf g_n}\eqno(1)$$ where 
$a_o$ is a new physical parameter with
units of acceleration and $\mu(x)$ is a function that is unspecified
but must have the asymptotic form $\mu(x) = x$ when $x<<1$ and 
$\mu(x) = 1$ where $x>>1$. 
\begin{figure}
\begin{center}
\includegraphics[height=10cm]{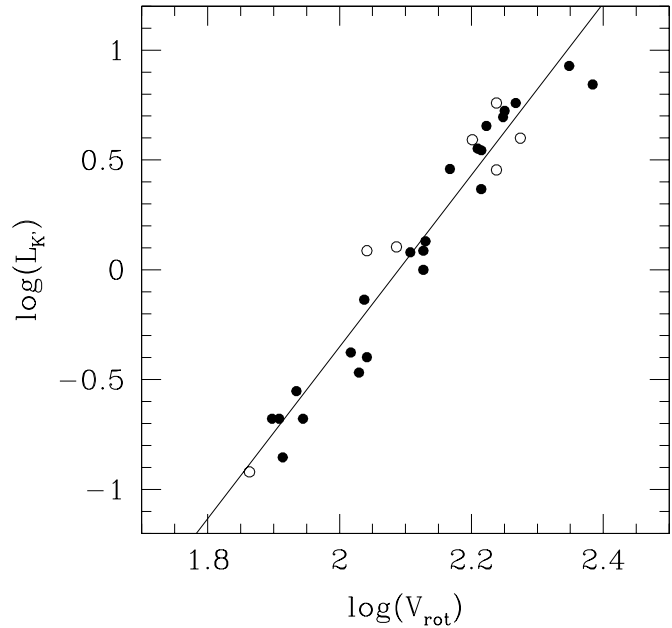}
\caption{The near-infrared Tully-Fisher relation of Ursa Major spirals
\cite{sv98}. 
The rotation velocity is the asymptotically constant value.
The line is a least-square fit to the data and has a slope of $3.9\pm 0.2$}
\end{center}
\end{figure}

The immediate consequence of this is that, in the limit of low 
accelerations, ${\rm g}=\sqrt{{\rm g_n}a_o}$.
For a point mass M, if we set g equal
to the centripetal acceleration $v^2/r$, then the circular velocity is
$$v^4 = GMa_o\eqno(2)$$ in the low acceleration regime.  
So all rotation curves are
asymptotically flat and there is a mass-velocity relation of the form
$M\propto v^4$.  These are aspects that are built into MOND so they
cannot rightly be called predictions.  However,
in the context of MOND, the  aspect of an asymptotically flat rotation curve 
is absolute.  Unambiguous examples of rotation curves 
(of isolated galaxies) that 
decline in a Keplerian fashion at a large distance
from the visible object would falsify the idea. 

The implied mass-rotation velocity relation explains a well-known
global scaling relation for spiral galaxies, the Tully-Fisher relation.
This is a correlation between the observed luminosity of spiral
galaxies and the characteristic rotation velocity, a relation of
the form $L\propto v^\alpha$ where $\alpha \approx 4$ if luminosity is
measured in the near-infrared.  If the mass-to-light ratio of galaxies
does not vary systematically with luminosity, then MOND explains this
scaling relation.
In addition, because it reflects underlying physical law, the
relation is as absolute.  The
TF relation should be the same for different classes of galaxies and the
logarithmic slope (at least of the MASS-velocity relation) must be 4.
Moreover, the relation is essentially one between the total
baryonic mass of a galaxy and the asymptotic flat rotational velocity--
not the peak rotation velocity but the velocity at large distance.
This is the most immediate prediction 
\cite{mcdb98,mgeal00}.  

The near-infrared TF relation for a sample of galaxies in the Ursa
Major cluster (and hence all at nearly the same distance)
is shown as a log-log plot in Fig.\ 1 where the 
velocity is that of
the flat part of the rotation curve \cite{sv98}.  The scatter about the 
least-square fit line of slope $3.9 \pm 0.2$ is consistent with observational
uncertainties (i.e., no intrinsic scatter).

Given the mean M/L in a particular band ($\approx 1$ in the 
K' band), this observed TF relation (and eq.\ 2) tells us that
$a_o$ must be on the order of $10^{-8}$ cm/s$^2$.  
It was immediately noticed by Milgrom
that $a_o\approx cH_o$ to within a factor of 5 or 6.
This cosmic coincidence suggests that MOND, if it
is right, may reflect the effect of cosmology on local particle dynamics.

\subsection{A critical surface density}

It is evident that the surface density of a system $M/R^2$ is proportional
to the internal gravitational acceleration.
This means that the critical acceleration may be rewritten as a 
critical surface density:
$$\Sigma_m \approx a_o/G.\eqno(3)$$  If a system, such as a spiral galaxy has
a surface density of matter greater than $\Sigma_m$, then
the internal accelerations are greater than $a_o$, so the system is
in the Newtonian regime.  In systems with $\Sigma \geq \Sigma_m$ (high
surface brightness or HSB
galaxies) there should be a small discrepancy between
the visible and classical Newtonian dynamical mass within the optical disk. 
But in low surface brightness (LSB) 
galaxies ($\Sigma<<\Sigma_m$) there is a low internal 
acceleration, so the discrepancy between the visible
and dynamical mass would be large.  
By this argument Milgrom predicted, before the actual
discovery of a large population of LSB galaxies, that there should be a 
serious discrepancy between the observable and dynamical mass within the
luminous disk of such systems-- should they exist.  They do exist, and
this prediction has been verified \cite{mcdb98}. 
\begin{figure}
\begin{center}
\includegraphics[height=10cm]{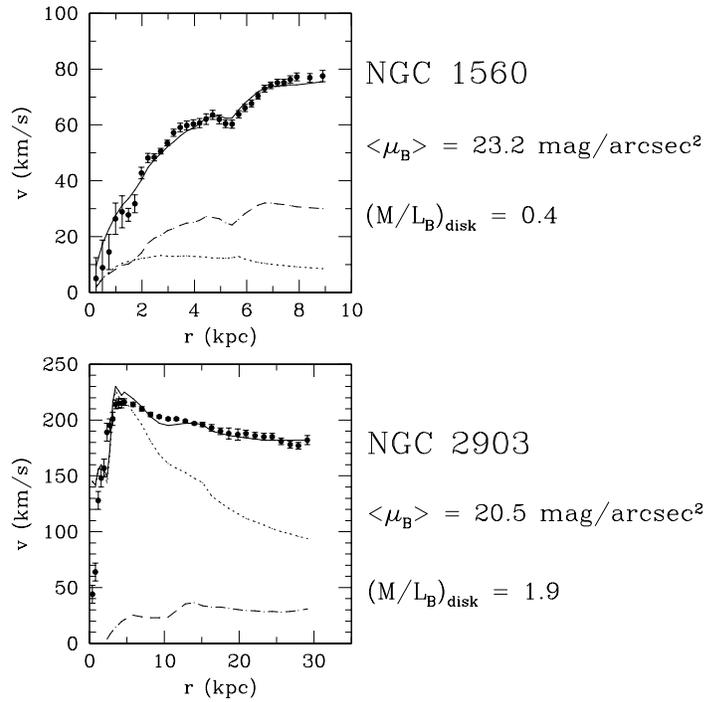}
\caption{The points show the observed 21 cm line rotation curves of
a low surface brightness galaxy, NGC 1560 and a high
surface brightness galaxy, NGC 2903.  The dotted and dashed
lines are the Newtonian rotation curves of the visible and gaseous
components of the disk and the solid line is the MOND rotation curve 
with $a_o=1.2\times 10^{-8}$ cm/s$^2$-- the value derived from the
rotation curves of 10 nearby galaxies \cite{BBS}.  Here
the only free parameter is the mass-to-light ratio of the visible 
component.}
\end{center}
\end{figure}

Moreover, spiral galaxies with a mean surface density near this limit -- 
HSB galaxies-- would be, within
the optical disk, in the Newtonian regime.  So one would expect that
the rotation curve would decline in a near Keplerian fashion to the 
asymptotic constant value.  In LSB galaxies, with mean surface density
below $\Sigma_m$, the prediction is that rotation curves would rise to
the final asymptotic flat value.  So there should be a general difference
in rotation curve shapes between LSB and HSB galaxies.
In Fig.\ 2 I show the observed rotation curves (points) of two galaxies, 
a LSB and HSB \cite{BBS}, where we
see exactly this trend.  This general effect in observed rotation curves
was pointed out in ref. \cite{cvg91}.

It is well-known that rotationally supported Newtonian systems tend to be 
unstable to
global non-axisymmetric modes which lead to bar formation and rapid
heating of the system \cite{ospe73}.  
In the context of MOND, these systems would be
those with $\Sigma > \Sigma_m$, so this would suggest that $\Sigma_m$
should appear as an upper limit on the surface density of rotationally 
supported systems.  This critical surface density is
0.2 g/cm$^2$ or 860 M$_\odot$/pc$^2$.  A more appropriate value of the mean
surface density within an effective radius would be $\Sigma_m/2\pi$
or 140 M$_\odot/pc^2$, and, taking
$M/L_b \approx 2$, this would correspond to a surface brightness of
about 22 mag/arc sec$^2$.  There is 
such an observed upper limit on the mean surface brightness of spiral 
galaxies and this is known as
Freeman's law \cite{free74}. The existence of such a limit
becomes understandable in the context of MOND.

\subsection{Pressure-supported systems}

Of course, spiral galaxies are rotationally supported.  But there other
galaxies, elliptical galaxies, which are
pressure supported-- i.e., they are held up against gravity by the random
motion of the stars.  There are numerous other examples of pressure-supported
systems such as globular clusters and clusters of galaxies, and often the
observable components of these systems have a velocity dispersion (or 
temperature) that does not vary much with position; i.e., they are near
``isothermal''.
With Newtonian dynamics, pressure-supported systems that 
are nearly isothermal have infinite extent.  But in the context of MOND
it is straightforward to demonstrate that such isothermal systems
are finite with the density at large radii falling roughly like $1/r^4$
\cite{ml84}.  

The equation of hydrostatic equilibrium for an isotropic, isothermal
system reads $${\sigma_r}^2 {{d\rho}\over{dr}} = -\rho g\eqno(4)$$
where, in the limit of low accelerations $g=\sqrt{GMa_o}/r$.
Here $\sigma_r$ is the radial velocity dispersion and $\rho$ is
the mass density.
It then follows immediately that, in this MOND limit,
$$\sigma_r^4 = {GMa_o\Bigl({{d\,ln(\rho)}\over{d\,ln(r)}}\Bigl)^{-2}}.
\eqno(5)$$
Thus, there exists a mass-velocity dispersion
relation of the form $$(M/10^{11}M_\odot) \approx (\sigma_r/100\,\,
kms^{-1})^4$$ which is similar to the observed
Faber-Jackson relation (luminosity-velocity dispersion relation) 
for elliptical galaxies \cite{fj}.   
This means that a MOND near-isothermal sphere 
with a velocity dispersion on the order of 100 km/s 
will always have a galactic mass.  This is not 
true of Newtonian pressure-supported objects.  Because of the appearance
of an additional dimensional constant, $a_o$, in the structure equation 
(eq.\ 4), MOND systems are much more constrained than their Newtonian
counterparts. 
 
Any isolated system which is nearly isothermal will 
be a MOND object.  That is because a Newtonian isothermal system  
(with large internal accelerations) 
is an object of infinite size and will always extend to the
region of low accelerations ($<a_o$).  At that point (${r_e}^2\approx 
GM/a_o$), MOND intervenes and the system will be truncated.
This means that the internal acceleration of any isolated 
isothermal system (${\sigma_r}^2/r_e$) is expected to be on the
order of or less than $a_o$ and that the mean surface density 
within $r_e$ will typically be $\Sigma_m$ or less 
(there are low-density solutions for MOND isothermal spheres,
$\rho<<{a_o}^2/G\sigma^2$, with internal accelerations less than $a_o$).  
It was pointed out long ago that elliptical galaxies
do appear to have a characteristic surface brightness \cite{fish}.
But the above arguments imply that the same should be true of any
pressure supported, near-isothermal system, from globular clusters
to clusters of galaxies.  Moreover, the same $M-\sigma$ relation 
(eq.\ 5) should apply to all such systems, albeit with considerable
scatter due to deviations from a strictly isotropic, isothermal
velocity field \cite{rhs00}. 

Most luminous elliptical galaxies are high surface brightness objects
which would imply a surface density greater than the MOND limit.  This
suggests that luminous elliptical galaxies should be essentially 
Newtonian objects, and, viewed in the traditional way, should evidence
little need for dark matter within the effective (or half-light) radius.
This does seem to be the case as demonstrated by dynamical studies
using planetary nebulae as kinematic tracers \cite{romeal,milsan}.

\subsection{Rotation curves of spiral galaxies}

Perhaps the most impressive observational success of MOND is the prediction
of the form of galaxy rotation curves from the observed distribution
of baryonic matter, stars and gas.  Basically, one takes the mean radial
distribution of light in a spiral galaxy as a precise tracer of the
luminous mass, includes the observed radial dependence of neutral
hydrogen (increased by 30\% to account for the primordial helium) and
assumes all of this is in a thin disk (with the occasional exception of
a central bulge component). One then solves the standard Poisson equation
to determine the Newtonian force, applies the MOND formula (eq.\ 1 
with a fixed value of $a_0$) to
determine the true gravitational force and calculates the predicted rotation
curve.  The mass-to-light ratio of the visible component is adjusted to
achieve the best fit to the observed rotation curve.  

The results are
spectacular considering that this is a one-parameter fit.
The solid curves in Fig.\ 2 are the results of such a procedure applied to
a LSB and HSB galaxy; this has been done for about 100 galaxies.
The fitted M/L values are not only reasonable, but demonstrate the
same trend with colour that is implied by population synthesis models
as we see in Fig. 3 \cite{sv98,bdj01}.
\begin{figure}
\begin{center}
\includegraphics[height=10cm]{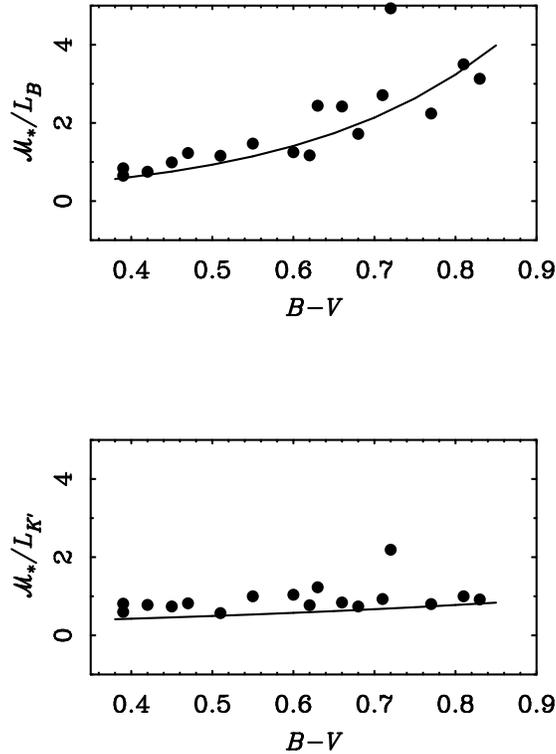}
\caption{MOND fitted mass-to-light ratios for the UMa spirals \cite{sv98}
in the B-band (top) and the K'-band (bottom)
plotted against B-V (blue minus visual) colour index.
The solid lines show predictions from
populations synthesis models \cite{bdj01}}
\end{center}
\end{figure}

Here I wish to emphasise another observed aspect of galaxy rotation curves--
a point that has been made, in particular, by Sancisi \cite{rsanc}.
For many objects, the detailed rotation curve appears to be extremely
sensitive to the distribution observable matter, even in LSB galaxies
where, in the standard interpretation, dark matter overwhelmingly
dominates within the optical image.  There are numerous examples of this--
for example, the LSB galaxy shown in Fig.\ 2 where we see that the
total rotation curve reflects the Newtonian rotation curve of the
gaseous component in detail.  Another example \cite{rsanc,zweal} is the
dwarf galaxy, NGC 3657.  Fig.\ 4 shows the surface densities of
the baryonic components, stars and gas, compared to the observed
rotation curve.  Again the dotted and dashed curves are the 
Newtonian rotation curves of the stellar and gaseous components and the
solid curve is the resulting MOND rotation curve.  The agreement with
observations is obvious.  

\begin{figure}
\begin{center}
\includegraphics[height=10cm]{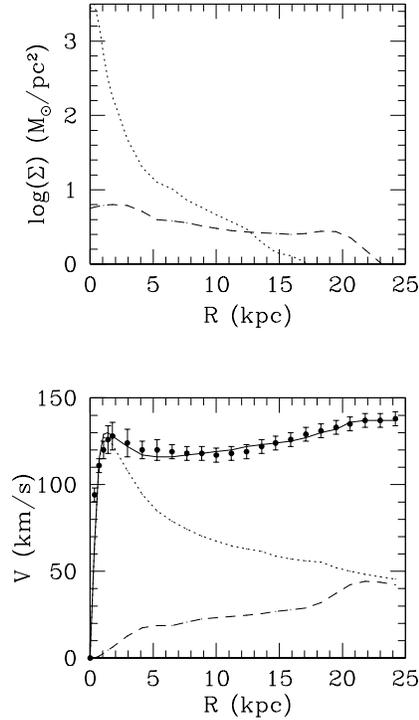}
\caption{The upper panel is the logarithm of the surface density of
the gaseous and stellar components of NGC 3657.  The lower panel shows
the observed rotation curve (points), the Newtonian rotation curves for
the stellar (dashed) and gaseous (dotted) components as well as the MOND
rotation curve (solid) \cite{rsanc,zweal}}
\end{center}
\end{figure}

For this
galaxy, there is evidence from the rotation curve of a central cusp in
the density distribution-- and, indeed, the cusp is seen in the light
distribution.  In cases where there is no conspicuous cusp in the light
distribution, there is no kinematic evidence for a cusp in the rotation
curve.  This would appear to make the entire discussion about 
cusps in halos somewhat irrelevant.  But equally striking in this case
is the gradual rise in the rotation curve at large radii.  This rise
is clearly related to the increasing dominance of the gaseous component
in the outer regions.  The point is clear: the rotation curve reflects
the global distribution of baryonic matter, even in the presence of a 
large discrepancy between the visible and Newtonian dynamical mass.  
This is entirely understandable (and predicted) in the context of modified
gravity in the form of MOND (what you see is all there is),
but remains mysterious in the context of dark matter

\subsection{Clusters of galaxies:  a phenomenological problem for MOND?}

It has been known for 70 years \cite{zwk33} that clusters
of galaxies exhibit a significant discrepancy between the Newtonian
dynamical mass and the observable mass,
although the subsequent discovery of hot X-ray emitting gas 
goes some way in alleviating the original discrepancy.  For an isothermal
sphere of hot gas at temperature T, the Newtonian dynamical 
mass within radius
$r_o$, calculated from the equation of hydrostatic equilibrium, is 
$$M_n = {{r_o}\over G} {{kT}\over m} 
\Bigl({{d\,ln(\rho)}\over{d\,ln(r)}}\Bigl),\eqno(6)$$
where $m$ is the mean atomic mass and the logarithmic density gradient
is evaluated at $r_o$.  This dynamical mass  
turns out to be typically about a factor of 4 or 5 larger than
the observed mass in hot gas and in the stellar content of the galaxies
(see Fig.\ 5, left \cite{rhs99}).

\begin{figure}
\begin{center}
\includegraphics[height=6cm]{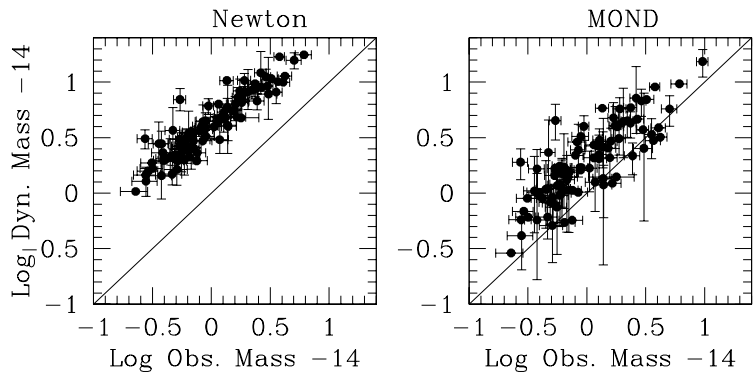}
\caption{({\it Left}) the Newtonian dynamical mass of clusters of 
galaxies within an observed cutoff radius ($r_{out}$) vs. the total 
observable mass in 93
X-ray emitting clusters of galaxies.
The solid line corresponds to $M_{dyn} = M_{obs}$ (no discrepancy). 
({\it Right}) the MOND dynamical mass 
within $r_{out}$ vs. the total observable mass for the same X-ray emitting 
clusters \cite{rhs99}}
\end{center}
\end{figure}

With MOND, the dynamical mass (eq.\ 5) is
given by $$M_m = {(Ga_o)}^{-1} {\Bigl({{kT}\over m}\Bigr)^2}
\Bigl({{d\,ln(\rho)}\over{d\,ln(r)}}\Bigl)^2,\eqno(7)$$ and,
using the same value of $a_o$ determined from nearby galaxy 
rotation curves, turns out to be, on average, a factor of two
larger than the observed mass (Fig.\ 5, right).  The discrepancy
is reduced but still present.   
This could be interpreted as a failure \cite{ageal02}, 
or one could say that MOND predicts that
the mass budget of clusters is not yet complete and that there is
more mass to be detected \cite{rhs99}.  The cluster missing mass
could, for example, be in neutrinos of mass 1.5 to 2 eV \cite{rhs03},
or in ``soft bosons'' with a large de Broglie wavelength \cite{rhs05}, 
or simply in heretofore undetected baryonic matter.
It would have certainly been a falsification of
MOND had the predicted mass turned out to be typically {\it less} 
than the observed mass in hot gas and stars.

\section{Relativistic MOND} 

MOND not only allows the
form of rotation curves to be precisely predicted from the 
distribution of observable matter, but it also explains certain systematic
aspects of the photometry and kinematics of galaxies and clusters:  the
presence of a preferred surface density in spiral galaxies and ellipticals--
the so-called Freeman and Fish laws; the
fact that pressure-supported nearly isothermal systems ranging from
molecular clouds to clusters of galaxies are characterised
by a specific internal acceleration, $a_o$ \cite{sm02};
the existence of a TF relation with small scatter-- specifically a
correlation between the baryonic mass and the asymptotically flat 
rotation velocity of the form $v^4\propto M$;
the Faber-Jackson relation for ellipticals, and with more detailed 
modelling, the Fundamental Plane \cite{rhs00};  
not only the magnitude of the discrepancy
in clusters of galaxies but also the fact that mass-velocity dispersion 
relation which applies to elliptical galaxies (eq.\ 5) extends to clusters
(the mass-temperature relation).  And it accomplishes all of this with a
single new parameter with units of acceleration-- a parameter determined from
galaxy rotation curves which is within an order of magnitude of the
cosmologically significant value of $cH_o$.  This is why several of us believe
that, on an epistemological level, MOND is more successful than dark matter.  
Further, many of these systematic aspects of
bound systems do not have any obvious connection to what has been
traditionally called the ``dark matter problem''.  This capacity to
connect seemingly unrelated points is the hallmark of a good theory.    
However, as I argued in the Introduction,
MOND will never be entirely credible to most astronomers and physicists 
until it makes some contact with more familiar physics-- until there
is an underlying and understandable physical mechanism for MOND
phenomenology. Below I consider that mechanism in terms of possible 
modifications of the theory of gravity.

\subsection{Steps to TeVeS}

TeVeS (tensor-vector-scalar) theory \cite{jdb04} is a relativistic
theory yielding MOND phenomenology in the appropriate limit.  Of
course, I do not need to belabour the advantages of a relativistic 
theory.  It allows one to address a number of issues on which MOND
is silent:  gravitational lensing, cosmology, structure formation,
anisotropies in the CMB.  The theory is complicated-- considerably
more complicated than GR-- in that involves additional dynamical
elements and is characterised by three additional free parameters
and a free function-- i.e., a function that is not specified by
any a priori considerations but may be adjusted to achieved the
desired result.  In this sense, TeVeS, like MOND itself, is a 
phenomenologically driven theory.  It is entirely ``bottom-up'' and
thereby differs from what is normally done in gravity theory
or cosmology.

As the name implies it is a multi-field theory; i.e., there are fields 
present other than the usual tensor field $g_{\mu\nu}$ of GR. It appears
that any viable theory of MOND as a modification of gravity must be
a multi-field theory; no theory based upon a single metric field can
work \cite{souwood}.  In TeVeS, the MOND
phenomenology appears as a ``fifth force'' mediated by a scalar field.
This fifth force must be designed to fall as $1/r$ and dominate over
the usual Newtonian force when the total force is below $a_0$ as
shown in Fig.\ 6.
\begin{figure}
\begin{center}
\includegraphics[height=8cm]{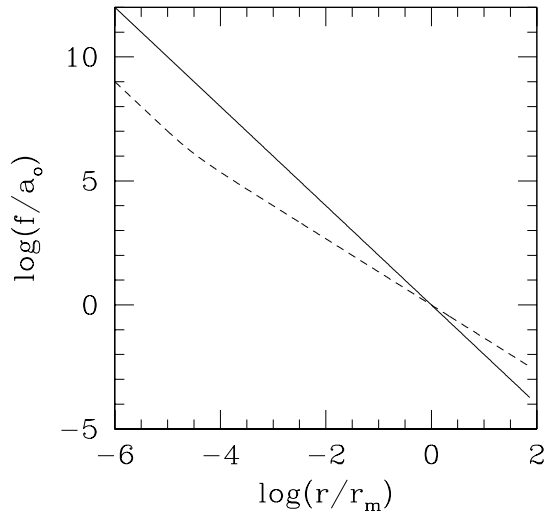}
\caption{MOND phenomenology as a result of multi-field modifications of
gravity.  The dashed curve shows the log force resulting from a scalar field
with a non-standard Lagrangian as a function of log radius in units
of the MOND radius $r_M = \sqrt{GM/a_0}$.  The solid line is the usual
Einstein-Newton force.}
\end{center}
\end{figure}

Now if we are proposing a fifth force, then that implies non-geodesic
motion and one may naturally ask about the validity of
the equivalence principle, even in its weak form expressing the 
universality of free fall (there are strong experimental constraints
on the composition independence of acceleration in a gravitational field).
The weak version of the equivalence principle can be preserved if there
is a specific form of coupling between the scalar field a matter--
one in which the scalar couples to matter jointly with the gravitational
or Einstein metric. This allows for the definition of a physical
metric, $\tilde{g}_{\mu\nu}$ that is distinct from the Einstein metric.
In the simplest sort of joint coupling the physical metric is 
{\it conformally} related to the Einstein metric, i.e.,
$$\tilde{g}_{\mu\nu} = f(\phi) g_{\mu\nu}.\eqno(8)$$
This is the case in traditional scalar-tensor theories such as the 
Brans-Dicke theory \cite{bd61}.  So the theory remains a metric
theory, but particle
and photons follow geodesics of the physical metric and not the
Einstein metric.  Of course, a great part of the beauty of GR is
that the gravitational metric is the metric of a 4-D space time
with Lorentzian signature-- gravitational geometry {\it is}
physical geometry.  It is beautiful, but the world doesn't have to be 
that way.

Another ingredient is necessary if the scalar field is to produce
MOND phenomenology.  In standard scalar-tensor theory, the 
scalar field Lagrangian is
$$L_s ={1\over 2} \phi_{,\alpha}\phi^{,\alpha}.\eqno(9)$$
Forming the action from this Lagrangian (and the joint coupling
with $g_{\mu\nu}$ to matter) and taking the condition
of stationary action leads, in the weak field limit, to the usual 
Poisson equation for $\phi$.  In other words,
the scalar force about a point mass falls as $1/r^2$ as in Brans-Dicke
theory.  Therefore, MOND requires a non-standard scalar field Lagrangian;
for example, something like
$$L_s = {1\over {2l^2}}F(l^2\phi_{,\alpha}\phi^{,\alpha}) \eqno(10)$$
where $F(X)$ is an, as yet, unspecified function of the usual scalar
invariant and $l$ is a length scale on the order of the
present Hubble scale ($\approx c/H_0$).  Bekenstein refers to this
as aquadratic Lagrangian theory or AQUAL.  
The condition of stationary action then leads to a scalar
field equation that, in the weak field limit, is
$$\nabla\cdot[\mu(|\nabla\phi|/a_0)\nabla\phi] = 4\pi G\rho \eqno(11)$$
where $a_0=c^2/l$ and $\mu = dF(X)/dX$.
This we recognise as the Bekenstein-Milgrom field equation \cite{bm84}
which produces MOND like phenomenology if   
$\mu(y) = y$ when $y<1$ or $F(X) = {2\over 3}X^{3\over 2}$.
Here, however, we should recall that $\phi$ is not the total gravitational
field but on the scalar component of a two-field theory.
Another phenomenological requirement on the free function is that
$F(X) \rightarrow \omega X$ in the limit where $X>>1$ (or $\nabla\phi>a_0$).
That is to say, the scalar field Lagrangian becomes standard in the
limit of large field gradients; the theory becomes equivalent to
Brans-Dicke theory in this limit.  This guarantees precise $1/r^2$
attraction in the inner solar system, but, to be consistent with
post-Newtonian constraints, it is necessary that $\omega>10^4$.

Looking at the form of $F$ required for MOND phenomenology, we see an
immediate problem with respect to cosmology.  In the limit of a 
homogeneous Universe, where $\nabla\phi\rightarrow 0$ and the 
cosmic time derivative, $\dot{\phi}$, dominates the invariant, i.e., $X<0$.
This means that the form of the free function must change in this limit
(this is a problem which persists in TeVeS).  But there is another more
pressing problem which was immediately noticed by Bekenstein and Milgrom.
In the MOND limit, small disturbances in the scalar field, scalar waves,
propagate {\it acausally}; i.e., $V_s = \sqrt{2} c$ in directions 
parallel to $\nabla\phi$.  This is unacceptable; a physically viable 
theory should avoid the paradoxes resulting from acausal propagation.

The superluminal propagation (or tachyon) problem led Bekenstein to
propose a second non-standard scalar-tensor theory for MOND--
phase-coupling gravitation or PCG \cite{jdb88a}.  Here, the scalar field
is taken to be complex, $\chi = q e^{i\phi}$ with the standard Lagrangian,
$$L_S = {1\over 2}[q_{,\alpha}q^{,\alpha} + q^2\phi_{,\alpha}\phi^{,\alpha}
 + 2V(q)]\eqno(12)$$
where $V(q)$ is the potential function of the scalar field.
The non-standard aspect is that only the phase couples to matter in the
usual conformal way, 
$$\tilde{g}_{\mu\nu} = e^{-\eta\phi}g_{\mu\nu}.\eqno(13)$$
This leads (weak field limit) to the field equation,
$$\nabla\cdot[q^2\nabla\phi] = {{8\pi G \rho}\over {c^2}}. \eqno(14)$$
So now we see that $q^2$ replaces the usual MOND interpolating function
$\mu$, but now $q$ is given by a second scalar field equation,
$${q^{,\alpha}}_{;\alpha} = q\phi_{,\alpha}\phi^{,\alpha}+V'(q).\eqno(15)$$ 
That is to say, the relation between $q^2$ and $\nabla\phi$ is now 
differential and not algebraic as in AQUAL theory.
Bekenstein demonstrated that if $V(q) = -Aq^6$ (a negative sextic
potential) then the predicted phenomenology is basically that of MOND
on a galactic scale.

Obviously the property $dV/dq<0$ cannot apply for all $q$ because
this would lead to instability of the vacuum, but there is a more
serious problem:  By a suitable redefinition of the fields,
it may be shown that, in the limit of very weak coupling
($\eta<<1$) the term on the left-hand side of eq.\ 15 may be
neglected-- that is to say, we are left with only the right-hand
side and the relation between $q^2$ and $\nabla\phi$
once again becomes algebraic as in AQUAL.
In other words, PCG approaches AQUAL in the limit of very weak coupling.
This suggests that PCG may suffer from a similar ailment as AQUAL;
indeed, there is a problem, but it appears as the absence of 
a stable background solution rather than superluminal propagation
\cite{jdb90}. But I only mention this because I want to emphasise that 
the weak coupling limit
of PCG is equivalent to the aquadratic theory; this turns out
to be a significant aspect of TeVeS.

At about the same time it was realized that there is a serious
phenomenological
problem with AQUAL or PCG or any scalar-tensor theory in which the
the relation between the physical and gravitational metrics is
conformal as in eqs.\ 8 or 13. That is, such a theory would predict
no enhanced deflection of photons due to the presence of the scalar field
\cite{bs94}.
Recall that photons and other relativistic particles follow null 
geodesics of the physical metric.  These are given by the condition
that $${d\tilde{\tau}}^2 = -{\tilde{g}_{\mu\nu}}dx^\mu dx^\nu = 0.
\eqno(16)$$ Now given the conformal relation between the two metrics
(eq.\ 8) you don't have to be a mathematical genius to see that
$d\tilde{\tau}=0$ corresponds to $d\tau=0$; i.e., null geodesics
of the two metrics coincide which means that photons also follow
geodesics of the gravitational metric where the scalar field doesn't
enter (except very weakly as an additional source).  Hence the scalar
field does not influence the motion of photons!

This has a major observational consequence: It would imply that, for
a massive cluster of galaxies, the Newtonian mass one would determine from the
kinematics of galaxies (non-relativistic particles) via the virial theorem
should be much greater than the mass one would determine from gravitational
deflection of photons (relativistic particles).  This is, emphatically, not
the case \cite{bs94}.  The lensing contradiction is a severe blow to 
scalar-tensor theories of MOND, at least for those with a conformal
coupling.

An obvious solution to this problem is to consider a non-conformal
relationship between the Einstein and physical metrics, for
example
$$\tilde{g}_{\mu\nu} = g_{\mu\nu}e^{-\eta\phi}-(e^{\eta\phi}
-e^{-\eta\phi})A_\mu A_\nu \eqno(17)$$
where now $A^\mu$ is a normalized vector field, i.e., $A_\mu A^\mu = -1$
\cite{ni72}.  
Basically, the conformal relation transforms the gravitational geometry by
stretching or contracting the 4-D space isotropically but in
a space-time dependent way.  This {\it disformal} transformation, eq.\ 17, 
picks out
certain directions for additional stretching or contracting.  Because we
would like space in the cosmological frame to be isotropic (the Cosmological
Principle) we should somehow arrange for the vector to point in the time
direction in the cosmological frame, which then becomes a preferred frame.
In the spirit of the ancient stratified theories \cite{ni72}, one may
propose an a priori non-dynamical vector field postulated
to have this property.  This may be combined with an 
AQUAL theory to 
provide MOND phenomenology with enhanced gravitational lensing \cite{rhs97}; 
in fact,
with the particular transformation given by eq.\ 17 one can show that the
relation between the total weak field force and the deflection of photons
is the same as it is in GR.  Hence relativistic and non-relativistic 
particles would both feel the same weak-field force.

The problem with this initial theory is that the non-dynamical vector 
field quite explicitly violates the principle of General Covariance
making it impossible to define a conserved energy-momentum tensor (this
has been known for some time \cite{lln74}).  This problem led Bekenstein to
endow the vector field with its own dynamics, and, hence, to 
TeVeS.

\subsection{The structure of TeVeS}

As the name implies, the theory is built from three fields.

a) The tensor:  This is the usual Einstein metric that we are all 
familiar with.  It's dynamics are given by the standard Einstein-Hilbert
action of GR:
$$S_T = {1\over{16\pi G}}\int{R\sqrt{-g}d^4x}.\eqno(18)$$
It is necessary that the tensor should be the Einstein metric because
we want the theory to approach GR quite precisely in the appropriate 
strong field limits.

b) The scalar:  We want the scalar, $\phi$, to provide a long-range
fifth force in the limit of low field gradients
Bekenstein takes the scalar field action to be
$$S_S = -{1\over{16\pi G}}\int{[{1\over 2}q^2 h^{\alpha\beta}
\phi_{,\alpha}\phi^{,\beta}+l^{-2}V(q)]\sqrt{-g}
d^4x}.\eqno(19)$$
Here I have kept the notation of PCG because the action is, in fact, the
weak coupling, or AQUAL limit, of PCG where there is no explicit 
kinetic term for the field $q$.  In other words, $q$ behaves as 
a non-dynamical auxiliary field where $q^2$ will play the role of 
$\mu$ in the Bekenstein-Milgrom field equation (the fact that this field
is non-dynamical does not violate General Covariance because it does not
act directly upon particles).  I use this bi-scalar notation because
I think it is important to realise that the auxiliary field could, in
fact, be dynamical.  This, in some respects, provides a plausible 
interpretation of the free function, $V(q)$, as a potential (let's call
it a pseudo-potential for now).  As we see
below, this can provide a basis for cosmological dark matter.

Another difference with standard 
scalar-tensor theory is that the invariant $h^{\alpha\beta}
\phi_{,\alpha}\phi_{,\beta}$ has replaced the usual scalar field invariant
$g^{\alpha\beta}\phi_{,\alpha}\phi_{,\beta}$ where
$$h^{\alpha\beta} = g^{\alpha\beta} - A^{\alpha}A^{\beta}\eqno(20)$$
and {\bf A} is the normalized vector field described below. Bekenstein
has shown that this simple replacement solves the superluminal propagation
problem of AQUAL theories of MOND.  The speed of scalar waves turns out to
be precisely $c$.  

c) The vector:  The dynamical normalized vector field
is necessary to provide the disformal transformation
and the enhanced gravitational lensing.  Bekenstein chose to describe
its dynamics through the action
$$S_V = {K\over{32\pi G}}\int{[F^{\mu\nu}F_{\mu\nu}-2({\lambda\over K})
(A^\mu A_\mu + 1)\sqrt{-g}d^4x}\eqno(21)$$
where $F_{\mu\nu}$ is the electromagnetic-like anti-symmetric tensor
constructed from {\bf A}
$$F_{\mu\nu} = A_{\nu ;\mu}-A_{\mu ;\nu}, \eqno(22)$$
and $\lambda$ is a Lagrangian multiplier function which enforces the
normalisation condition $A_\mu A^\mu = -1$.  K is a new parameter which 
determines the strength of the vector field coupling.

All of this is combined with the particle action
$$S_P = -mc \int (-\tilde{g}_{\mu\nu} {{dx^\mu}\over {dp}}
{{dx^\nu}\over{dp}})^{1\over 2} dp \eqno(23)$$
where $\tilde{g}_{\mu\nu}$ is the physical metric disformally
related to the Einstein metric as in eq.\ 17.  This guarantees that the
deflection of photons is given by
$$\delta\theta = {2\over{c^2}}\int{f_\perp dl} \eqno(24)$$
where the integral is over the line-of-sight and $f_\perp$ is the perpendicular
component of the total weak-field force, Newtonian and scalar.

The free parameters of the theory are $\eta$, the scalar field coupling,
$K$ the vector field coupling, and $l$ the characteristic length scale
determining the MOND acceleration scale ($a_0=c^2/l$).   
It can be shown that, as the parameters $\eta$ and $K$ approach zero,
the theory reduces to GR, as it should do.  The free
function is $V(q)$ or the pseudo-potential of the auxiliary $q$ field.
I could have absorbed the length scale $l$ into $V(q)$ but, following
Bekenstein, I 
choose to express it explicitly in order to render $V(q)$ unitless.

In the weak-field static limit, the scalar field equation is of
the Bekenstein-Milgrom form:
$$\nabla\cdot(\mu {\bf f_s}) = 4\pi G\rho \eqno(25)$$
where, in my notation, the scalar force is given by
$f_s = \eta c^2 \nabla\phi$ and $\mu = q^2/2\eta^2$.  Making
use of these expressions, we may show that the MOND interpolating 
function is then given by the algebraic relation,
$${{dV(\mu)}\over{d\mu}}= -{{{f_s}^2}\over {a_0}^2}= -X\eqno(26)$$
where $a_0 = c^2/l$.  This, of course, necessitates
$V'(\mu)<0$ in the static domain.

Now, to obtain MOND phenomenology, it
must be the case that $\mu(X) = \sqrt{X}$ in the low acceleration
limit.  For example, $V(\mu)=-{1\over 3} \mu^3$ would work
(recalling the relation between $\mu$ and $q$ above, we see that this
gives rise to the negative sextic potential in PCG).  But this leaves
us with the old problem of extending AQUAL into the cosmological 
regime where $X<1$.  
\begin{figure}
\begin{center}
\includegraphics[height=8cm]{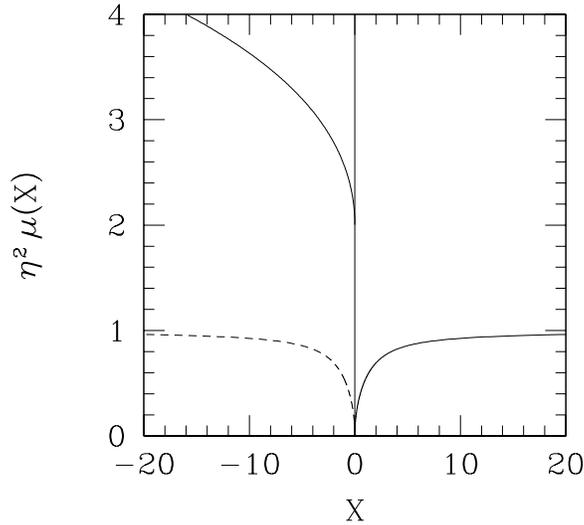}
\caption{Bekenstein's trial free function shown, $\mu(X)$ (solid
curve) where X is defined as $\eta^2l^2\phi_{,\alpha}\phi^{,\alpha}$.  
There are two discontinuous branches for cosmology
($X<0$) and for quasi-static mass concentrations ($X>0$).  The dashed
curve shows one possibility for avoiding the discontinuity (eq.\ 28).}
\end{center}
\end{figure}

Bekenstein chose to solve this problem by taking a free function that
provides two separate branches for $\mu(X)$-- one for static
mass concentrations, where the spatial gradients of $\phi$ dominate,
and one for the homogeneous evolving Universe where the temporal
derivative dominates. Specifically,
$$X = {1\over 4}\mu^2(\eta^2\mu-2)^2(1-\eta^2\mu)^{-1}\eqno(27)$$
($\eta^2$ appears because my definition of $\mu$ differs from Bekenstein's).
This two branch, $\mu(X)$, is shown in
Fig.\ 7, where now we are defining $X$ more generally as
$X = \eta^2 l^2\phi_{,\alpha}\phi^{,\alpha}$ 
The corresponding
pseudo-potential, $V(\mu)$, is shown in Fig.\ 8. 
\begin{figure}
\begin{center}
\includegraphics[height=8cm]{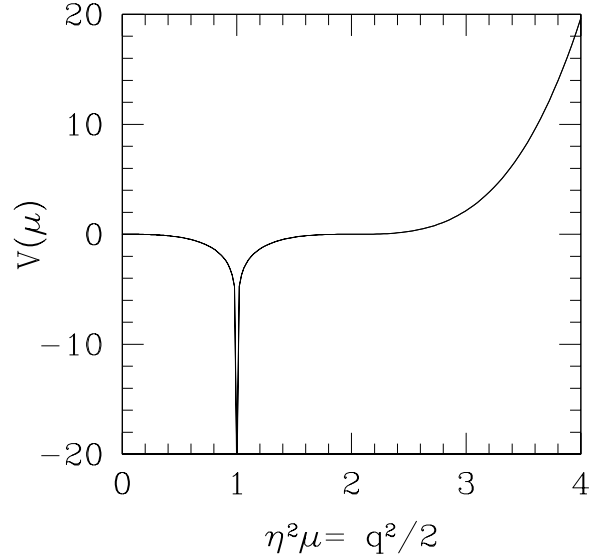}
\caption{The pseudo-potential ($V(\mu)$) corresponding to the $\mu(X)$ shown
in Fig.\  7}
\end{center}
\end{figure}

If we interpret $V(\mu)$ as the potential of an implicitly dynamical field, 
it is certainly a rather curious-looking one-- with the infinite pit at
$\eta^2\mu=1$.  It also illustrates the peculiar aspect of the two-branch
form of $\mu$.  For cosmological solutions, $\eta^2\mu = 2$ is an attractor;
i.e., the $\mu$ field seeks the point where $dV/d\mu = 0$ \cite{scdeal}.
However, on the outskirts of galaxies $\eta^2\mu\rightarrow 0$ as it must
to provide the $1/r$ scalar force.  So somehow, in progressing from
the galaxies to the cosmological background $\eta^2\mu$ must jump from
0 to 2 apparently discontinuously (photons propagating in a cosmological
background also have to make this leap).  This problem indicates that
such a two-branch $\mu(X)$ may not be appropriate, but more on this below.

\section{TeVeS: Successes, issues and modifications}

\subsection{Successes of TeVeS}
The theory is an important development because it solves several of 
the outstanding problems of earlier attempts.:

\medskip
\noindent 1.) While providing for MOND phenomenology in the form of the old
non-relativistic Bekenstein-Milgrom theory, it also allows for enhanced
gravitational lensing.  It does this in the context of a proper
covariant theory,  albeit by construction-- by taking the particular
disformal relation between the physical and gravitational metrics given
by eq.\ 17. This aspect of the theory has favourably tested on a sample of
observed strong lenses \cite{zheal}, although there are several case with
unreasonable implied mass-to-light ratios.

\medskip
\noindent 2.)  It has been shown \cite{jdb04,gian05} that, 
for TeVeS, the static post-Newtonian
effects are identical to those of GR; that is to say, the 
Eddington-Robertson post-Newtonian parameters are
$\gamma=\beta=1$ as in GR.  This provides consistency with a range
of Solar System gravity tests such as light deflection and radar
echo delay.

\medskip
\noindent 3.)  Scalar waves propagate causally ($v_s\leq c$).  
This is true because
the new scalar field invariant $h^{\alpha\beta}\phi_{,\alpha}
\phi_{,\beta}$ ($h^{\alpha\beta}$ is a new tensor built from the
Einstein metric and the vector field (eq.\ 20))
replaces the standard invariant in the scalar field
Lagrangian (eq.\ 9).  This is a major improvement over the old
AQUAL theory, but also one which relies upon the presence of the
vector field.

\medskip
\noindent 4.)  Gravitational waves propagate causally if $\phi>0$.  
One can show
\cite{jdb04} that the speed of the standard tensor waves is given by
$V_g = ce^{-\eta\phi}$.  This means that the cosmology must 
provide $\phi>0$ in a natural way.
Moreover, there is a prediction here which is possibly testable, and
that is $V_g<c$.  If an event, such as a gamma-ray burst, also produces
gravitational radiation (as is likely), the gravitational waves should
arrive somewhat later than the gamma rays.

\medskip
\noindent 5.)  The theory allows for standard FRW cosmology and, at least 
in the linear regime, for a MONDian calculation of structure formation
\cite{scdeal}.  Moreover, there is an evolving dark energy 
(quintessence) which is
coupled to the background baryon density, offering a possible 
solution to the near coincidence of these components at the present
epoch.  This comes about through the presence of
$V(\mu)$ as a negative pressure fluid in the Friedmann equations.  The 
cosmological value of the dark energy density, $V(\mu)$, corresponds
to the minimum of an effective potential
$V_{eff} = V(\mu) + B(\rho\tau)/\mu$ where $B$ is a function of the product
of cosmic time $\tau$ and the baryonic mass density $\rho$
(it is identical in this sense to PCG in a cosmological context
\cite{rhs89}).  

\subsection{Remaining issues}
In spite of these important successes there are a number of problems 
that the theory is yet to confront:

\medskip
\noindent 1.) The discontinuous 
$\mu(X)$.  The two discontinuous branches (Fig.\ 7)-- 
one for cosmology and one for quasi-static mass concentrations-- appears
awkward, particularly if the free-function is interpreted as a potential
of the $\mu$ field.  Moreover, this presents very practical problems for 
gravitational lensing and calculation of 
structure formation into the non-linear regime. 
But more seriously, it appears that such two branch $\mu$ 
may be an intrinsic
aspect of a theory with the structure of TeVeS.  One could propose
(as in \cite{zf05}) that the space-like branch of $\mu$ is simply 
reversed at the
at the $\mu=0$ axis (see dotted line in Fig.\ 7), so, instead of
eq.\ 27, Bekenstein's free
function could be expressed as 
$$X = \pm{{3\mu^2}\over {1-\eta^2\mu}}.\eqno(28)$$  
However, the pseudo-potential, $V(\mu)$, would
then also be double valued which would appear distinctly unphysical if
this is really to be identified with the potential of a implicitly
dynamical scalar $\mu$ (or $q$).  In my opinion, the only solution to
this problem is to alter the structure of the theory (see below).

\medskip
\noindent 2.) Even given
a $\mu(X)$ with two branches, the separation between
quasi-static and cosmological phenomena is artificial.  Eq.\ 26, which  
provides the relation between the scalar field gradient and $\mu$,
should also contain the cosmic time derivative of the scalar field 
because this is likely to
be of the same order as $dV/d\mu$; i.e., eq.\ 26 should read
$${{dV}\over{d\mu}}= -{{{f_s}^2}\over {{a_0}^2}} +{{\eta^2l^2\dot{\phi}^2}
\over{c^2}}.
\eqno(29)$$   Therefore the free function, relevant
to mass concentrations, may also be thought of as an evolving effective 
potential (this can actually be an advantage which I make use of
below).

\medskip
\noindent 3.)  This is a preferred frame theory that violates 
the Lorentz invariance
of gravitational phenomena.  This is because of the cosmic vector field
{\bf{A}}. In the cosmic frame, only the time component of {\bf{A}} is 
non-zero but for frames in relative motion with respect to the CMB
spatial components also develop non-zero values, and this has a 
real effect on particle dynamics.  In
the Solar System for example, there should be gravitational ether drift
effects, such as a polarisation of the earth-moon orbit along the direction
of $\bf{w}$, the
velocity vector with respect to the CMB.  Such effects, in conservative
theories, are quantified by two post-Newtonian parameters \cite{will},
$\alpha_1$ and $\alpha_2$, which enter the effective Lagrangian of an
N-body system as the coefficients of terms containing $v\cdot w/c^2$
where $\bf{v}$ is the velocity with respect to the centre-of-mass of the
N-body system.  These parameters are experimentally constrained; for example,
$\alpha_1<10^{-4}$ on the basis of Lunar Laser Ranging \cite{mnv96}.

It is important to determine predicted values of $\alpha_1$ and $\alpha_2$
for TeVeS.  A reasonable guess is that these post-Newtonian parameters will
approach zero as the free parameters of the theory, $\eta$ and $K$ approach
zero \cite{rhs06}.  That is because in this limit the theory approaches 
GR, and in GR there are no preferred frame effects.  Whether or not
the resulting constraints on $\eta$ and $K$ are consistent with other
aspects of Solar System and galaxy phenomenology remains to be seen.

\begin{figure}
\begin{center}
\includegraphics[height=8cm]{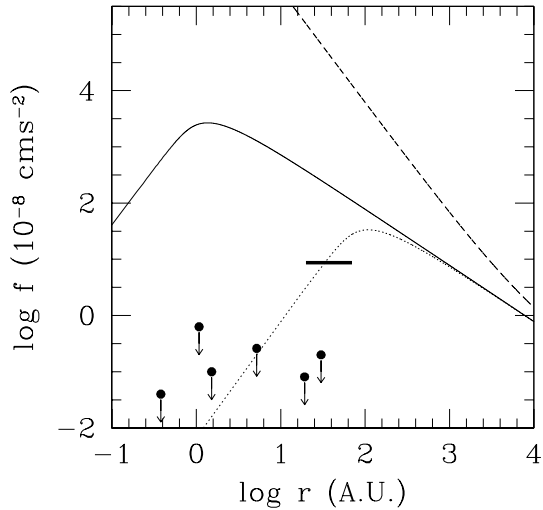}
\caption{The dashed curve is the log of the 
the total force ($f_t = f_s+f_N$),
in units of $10^-{8}$ cm/s$^2$ plotted against the log of
the radial distance from the sun in astronomical units
for TeVeS.  The dotted curve is the
anomalous force (the non-inverse square force) for Bekenstein's
initial choice of free function with $\eta =0.01$.  The long dashed
curve is the same but with $\eta=0.1$.  
Observed constraints on the non-inverse square part of the 
acceleration are (left to right): 
from the precession of perihelion of Mercury, and of Icarus, 
from variation of Kepler's constant between Earth and Mars, between 
inner planets and Jupiter, Uranus or Neptune, respectively.  
The horizontal bar is the Pioneer anomaly range. From reference \cite{bs05}.}
\end{center}
\end{figure}

\medskip
\noindent 4.)  In the outer solar system the 
force is not precisely inverse square.
For example, in the context of Bekenstein's free function, the 
non-inverse square component of the force is shown, as a function of
radius, in Fig.\ 9 for two different values of the scalar coupling 
strength, $\eta$.  Constraints from planetary motion are shown by the
upper limits \cite{rhs06}.  Such a deviation, at some level, is an aspect
of any multi-field theory of MOND \cite{rhs97}, 
and it may be a problem or
it may be a blessing.  A non-inverse square component of the force, in
the form of a constant acceleration, is indicated by Doppler ranging to
both the Pioneer spacecrafts (indicated by the horizontal bar in
Fig.\  8)\cite{ander}.  If this effect is confirmed, it would be a 
major discovery,
indicating that gravity is not what we think it is beyond the inner
solar system.

\medskip 
\noindent 5.) As 
I mentioned in the Introduction, there is compelling evidence
for cosmological dark matter-- a pressureless fluid which appears to
affect early large scale 
structure formation (evident in the CMB anisotropies) and
the more recent expansion history of the Universe (evident in the 
SNIa results).  The weight of this evidence implies that a 
proper theory of MOND should
at least simulate the cosmological effects of the apparent dark matter, 
again not an evident aspect of TeVeS.  

\medskip
In a general sense, the theory, at present, is intricate
and misses a certain conceptual simplicity.
There are several loose threads which
one might hope a theory of MOND to tie up.  For example, 
the MOND acceleration parameter, $a_0$, is put in by hand, as an 
effective length scale $l$; the observational fact that $a_0\approx cH_0$
remains coincidental.  This seems unfortunate because this coincidence
suggests that MOND results from the effect of cosmology on local 
particle dynamics, and, in the theory as it now stands, no such
connection is evident. 

Finally, by mentioning these problems, I do not wish to imply that
TeVeS is fundamentally flawed, but that it is not yet the theory in
final form.
In this procedure, building up from the bottom, the approach to
the final theory is incremental.

\subsection{Variations on a theme: biscalar-tensor-vector theory}

The motivation behind this variation is to use the basic elements of
TeVeS in order to construct a cosmologically effective theory of MOND.
The goals are to reconcile the galaxy scale success of MOND
with the cosmological evidence for CDM and to provide a cosmological basis
for $a_0$ \cite{rhs05}.  

There are two essential differences with TeVeS in original form:  First,
the auxiliary field $q$ is made explicitly dynamical as in PCG.
This is done by introducing a kinetic term for $q$
in the scalar action (eq.\ 19), i.e., $q_{,\alpha}q^{,\alpha}$.
Secondly, one makes use of the preferred frame to separate the spatial
and time derivatives of the matter coupling scalar field $\phi$ at
the level of the Lagrangian. Basically, this is done by defining new
scalar field invariants.  If we take the usual invariant to be 
$I=g^{\alpha\beta}\phi_{,\alpha}\phi_{,\beta}$ and define
$J= A^\alpha A^\beta\phi_{,\alpha}\phi_{,\beta}$, $K=J+I$,
then we can readily see that $J$ is just the square of the time derivative
in the preferred cosmological frame ($\dot{\phi}^2$) and $K$ is the 
spatial derivative squared in that frame ($\nabla\phi\cdot\nabla\phi$). 
The scalar field Lagrangian is then taken to be
$$L_s = {1\over 2}[q_{,\alpha}q^{,\alpha} + h(q)K - f(q)J + 2V(q)].
\eqno(30)$$

So, separate functions of $q$ multiply the spatial and temporal gradients
of $\phi$ in the cosmological frame.  This means that the potential for
$q$ becomes an effective potential involving the
cosmic time derivative, $\dot{\phi}$ for both the homogeneous cosmology
and for quasi-static mass concentrations.  Indeed, one can show, given
certain very general conditions on the free functions, $q$ at a large
distance from a mass concentration approaches its cosmological value.
There is smooth transition between mass concentrations and cosmology.
Moreover, if I take $h(q)\approx q^2$, $f(q)\approx q^6$ and a simple
quadratic bare potential $V(q)\approx Bq^2$, I obtain
a cosmological realisation of Bekenstein's PCG with a negative sextic
potential \cite{jdb88a} but where the coefficient in the potential, and
hence $a_0$, is identified with the cosmic $d\phi/dt$.

There are two additional advantages of making $q$ dynamical.
First of all, as the $q$ field settles to the evolving potential minimum,
oscillations of this field about that minimum inevitably develop.  
If the bare potential has a quadratic form, then these oscillations 
constitute CDM in the form of ``soft bosons'' \cite{prs}.  Depending upon 
the
parameters of the theory, the de Broglie wavelength of these bosons may
be so large that this dark matter does not cluster on the scale of galaxies
(but possibly on the scale of clusters).  A cosmological effective
theory of MOND produces cosmological CDM for free.

\begin{figure}
\begin{center}
\includegraphics[height=8cm]{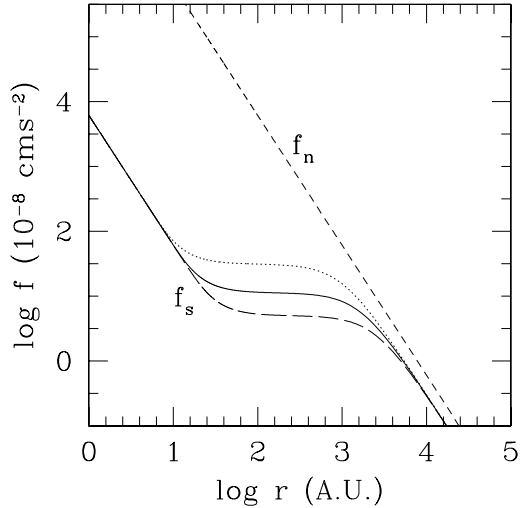}
\caption{The Newtonian (dashed curve) and scalar (solid curves) force in 
the Solar System in the context of the biscalar theory. The different 
curves correspond to different values of scalar coupling constant $\eta$.
This should be compared with Fig.\ 6 which shows the Newtonian and
scalar forces for TeVeS with the initial free function.}
\end{center}
\end{figure}

A second advantage is that appropriately chosen free functions
can reproduce the Pioneer anomaly in the outer Solar System-- both the
magnitude ($\approx 8\times 10^{-9}$ cm/s$^2$) and the form-- constant beyond
20 AU (see Fig.\ 10).  
It does this while being consistent with the form of galaxy
rotation curves \cite{rhs06}.  Of course the presence of three free functions
appear to give the theory considerable arbitrariness, but, in fact,
the form of these functions is strongly constrained by Solar System,
galaxy and cosmological phenomenology.

Many other modifications of TeVeS are possible.  For example, it may only
be necessary to make the auxiliary field $q$ explicitly dynamical and choose
a more appropriate form of the free function.  The number of 
alternative theories is likely to be severely restricted by the demands 
imposed by observations-- ranging from the solar system, to galaxies, to
clusters, to gravitational lensing, to cosmology.  The hope is that the
number of survivors is not less than one. 

\section{Conclusions}

Here I have outlined the attempts that have been made to
define modifications of gravity that may underly the highly successful
empirically-based MOND, proposed by Milgrom as an alternative to dark
matter in bound self-gravitating systems.  
These attempts lead inevitably to a multi-field theory of
gravity-- the Einstein metric to provide the phenomenology of GR in
the strong field limit, the scalar field to provide the MOND phenomenology
most apparent in the outskirts of galaxies and in low surface brightness
systems, and the vector field to provide
a disformal relation between the Einstein and physical metrics-- necessary
for the observed degree of gravitational lensing.  I re-emphasise that this
process has been entirely driven by phenomenology and the need to cure
perceived pathologies; there remains no connection to more a priori
theoretical considerations or grand unifying principles such as
General Covariance or Gauge Invariance.  It would, of course, be
a dramatic development if something like MOND were to emerge as a 
incidental consequence of string 
theory or a higher dimensional description of the Universe, but, 
in my opinion, 
this is unlikely.  It is more probable that an empirically based
prescription, such as MOND, will point the way to the correct theory.

The coincidence between the critical acceleration and $cH_0$ (or possibly
the cosmological constant) must be an essential clue.  MOND must be 
described by an effective theory; that is, the theory predicts
this phenomenology only in a cosmological context.  The aspect, and
apparent necessity, of a preferred frame invites further 
speculation:  Perhaps cosmology is described by a preferred frame
theory (there certainly is an observed preferred frame) with a long
range force mediated by a scalar field coupled to a dynamical vector field
as well as the gravitational metric.  With the sort of bi-scalar Lagrangian
implied by TeVeS, the scalar coupling to matter becomes very weak in
regions of high field gradients (near mass concentrations).  This 
protects the Solar System from detectable preferred frame
effects where the theory essentially reduces to General Relativity.
Because we live a region of high field gradients, 
we are fooled into
thinking that General Relativity is all there is.  Only the relatively
recent observations of the outskirts of galaxies or objects of low
surface brightness (or perhaps the Pioneer anomaly) reveal that there 
may be something more to gravity.

On the other hand, it may well be that we have been pursuing a mirage
with tensor-vector-scalar theories.  Perhaps the basis of MOND lies,
as Milgrom has argued, with modified particle action-- modified inertia--
rather than modified gravity \cite{m94a,m99}.  For a classical
relativist this distinction between modified gravity and modified inertia
is meaningless-- in relativity, inertia and gravity are two sides
of the same coin; one may be transformed into another by a change of frame.
But perhaps in the limit of low accelerations, lower than the fundamental
cosmological acceleration $cH_0$, that distinction is restored \cite{m05}.

It is provocative that the Unruh radiation experienced by a uniformly
accelerating observer, changes its character at accelerations below
$c\sqrt{\Lambda}$ in a de Sitter universe \cite{m99}.  
If the temperature difference
between the accelerating observer and the static observer in the de Sitter
Universe is proportional to inertia, then we derive an inertia-acceleration
relation very similar to that required by MOND \cite{m99}.
At present this is all very speculative, but it presents the possibility 
that we may
be going down a false path with attempted modifications of
GR through the addition of extra fields.

In any case, the essential significance of TeVeS is
not that it, at present,
constitutes the final theory of MOND. Rather, the theory provides a 
counter-example to the often heard
claim that MOND is not viable because it has no covariant basis. 

It is a pleasure to thank Jacob Bekenstein and Moti Milgrom for
helpful comments on this manuscript and for many enlightening conversations 
over the
years.  I also thank Renzo Sancisi for helpful discussions on the
``dark matter-visible matter coupling'' in galaxies and Martin Zwaan
for sending the data which allowed me to produce Fig.\ 4.
I am very grateful to the organisers of the Third Aegean Summer School
on the Invisible Universe, and especially, Lefteris Papantonopoulos, 
for all their efforts in making this school a most enjoyable and stimulating
event.


\begin{thebibliography}{0}
\bibitem [1]{speal03} D.N. Spergel, et al.: {\it Astrophys.J.Suppl.}, 
     148, 175 (2003) [astro-ph/0302209]
\bibitem [2] {pereal99} Perlmutter S, et al. 
    1999, \apj, 517, 565
\bibitem [3] {garn} P.M.\ Garnevitch et al.': \apj, 493, 53 (1998)
\bibitem [4] {tonry} J.L.\ Tonry et al.: \apj, 594, 1 (2003)
    [astro-ph/0305008]
\bibitem [5] {sanch} Sanchez, A.G. et al. \mnras, (in press, 2005)
\bibitem [6] {eisen} Eisenstein, D.J. et al. 2005, \apj, 633, 560
\bibitem [7] {apms} C.\ Armendariz-Picon, V.\ Mukhanov, P.J.\ Steinhardt:
   \pr, D63, 103510 (2001)
\bibitem [8] {cct03} S.\ Capozziello, S.\ Carloni, A. Troisi: 
astro-ph/0303041 (2003)
\bibitem [9]{ceal} S.M. Carroll, V. Duvvuri, M. Trodden, M.S. Turner:
    astro-ph/0306438 (2003)
\bibitem [10] {dvali} G.\ Dvali, G.\ Gabadadaze, M.\ Porrati:
    {\it Phys.Lett.}, B485, 208 (2000) [hep-th/0005016]
\bibitem [11]{zwk33} F.\ Zwicky: {\it Helv.Phys.Acta},
    6, 110 (1933)
\bibitem [12]{Bos78} A.\ Bosma {\it The Distribution and
    Kinematics of Neutral Hydrogen in Spiral Galaxies of Various
    Morphological Types}, PhD Dissertaion, Univ.
    of Groningen, The Netherlands (1978)
\bibitem [13]{bg89} K.G.\ Begeman: \aa 223, 4 (1989)
\bibitem [14]{bdsl83} J.R.\ Bond, A.S.\ Szalay: \apj, 274, 443 (1983)
\bibitem [15] {deluc} G.\ de Lucia et al.: \mnras,348, 333 (2004)
\bibitem [16] {debeal} W.J.G.\ de Blok, S.S.\ McGaugh, V.C.\ Rubin:
    \aj, 122, 2396 (2001) 
\bibitem [17]{m83a} M.\ Milgrom: \apj 270, 365 (1983a)
\bibitem [18]{m83b} M.\ Milgrom: \apj 270, 371 (1983b)
\bibitem [19]{m83c} M.\ Milgrom: \apj 270, 384 (1983c)
\bibitem [20] {wil} J.T.\ Wilson, ed.: {\it Continents Adrift and 
    Continents Aground} Scientific American, W. H. Freeman \& Company, 
    San Francisco (1976)
\bibitem [21] {sm02} R.H.\ Sanders, S.S.\ McGaugh:  {\it Ann.Rev.Astron.
   Astrophys.}, 40, 263 (2002)
\bibitem [22] {jdb04} J.D.\ Bekenstein, Phys.Rev.D,
    70, 083509 (astro-ph/0403694) (2004)
\bibitem [23]{mcdb98} S.S.\ McGaugh, W.J.G.\ de Blok:
     \apj, 499, 66 (1998)
\bibitem [24]{mgeal00} S.S.\ McGaugh, J.M.\ Schombert, G.D.\ Bothun, 
    W.J.G.\ de Blok: \apj, 533, L99 (2000) 
\bibitem [25]{sv98} R.H.\ Sanders, M.A.W.\ Verheijen:
    \apj, 503, 97 (1998)
\bibitem [26]{BBS} K.G.\ Begeman, 
    A.H.\ Broeils, R.H.\ Sanders: \mnras, 249, 523 (1991)
\bibitem [27]{cvg91} S.\ Casertano, 
    J.H. van Gorkom: \aj, 101, 1231 (1991)
\bibitem [28]{ospe73} J.P. Ostriker,
    P.J.E. Peebles: \apj 186, 467 (1973)
\bibitem [29] {free74} K.C.\ Freeman: \apj 160, 811 (1970)
\bibitem [30] {ml84} M.\ Milgrom: \apj 287, 571 (1984)
\bibitem [31]{fj} S.M.\ Faber, R.E.\ Jackson:
    \apj 204, 668 (1976)
\bibitem [32]{fish} R.A.\ Fish: \apj 139, 284 (1964)
\bibitem [33]{rhs00} R.H.\ Sanders \mnras, 313, 767 (2000) 
\bibitem [34] {romeal} A.J.\ Romanowsky et al.: {\it Science},
    301, 1696 (2003)
\bibitem [35] {milsan} M.\ Milgrom, R.H.\ Sanders: \apj,
    599, L25 (2003) 
\bibitem [36]{bdj01} E.F.\ Bell, R.S.\ de Jong:   
    \apj 550, 212 (2001)
\bibitem [37] {rsanc} R.\ Sancisi: IAU Symp. 220, Eds: S. D. Ryder, 
      D. J. Pisano, M. A. Walker, and K. C. Freeman. 
      San Francisco: ASP, p.233 (2004)
\bibitem [38] {zweal} M.A.\ Zwaan, J.M.\ van der Hulst, A.\ Bosma:
    (in preparation 2005) 
\bibitem [39] {rhs99} R.H.\ Sanders: \apj, 512, L23 (1999)
\bibitem [40] {ageal02} A.\ Aguirre, J.\ Schaye, E.\ Quataert: \apj,
    561, 550 (2002)
\bibitem [41] {rhs03} R.H.\ Sanders: \mnras, 342, 901 (2003)
\bibitem [42] {rhs05} R.H. Sanders: \mnras, 363, 459 (2005)
\bibitem [43] {souwood} M.E.\ Soussa, R.P.\ Woodard,  
    {\it Phys.Lett.} B578, 253 (2004) 
\bibitem [44]{bd61} C.\ Brans, R.H.\ Dicke, 
    \pr, 124, 925 (1961)
\bibitem [45]{bm84}J.D.\ Bekenstein,. 
    M.\ Milgrom: \apj, 286, 7 (1984) 
\bibitem [46]{jdb88a} J.D.\ Bekenstein: 
    {\it Second Canadian Conference on General Relativity
    and Relativistic Astrophysics}, eds. Coley, A., Dyer, C., Tupper, T.,
     p.\ 68 . Singapore: World Scientific (1988)
\bibitem [47]{jdb90} J.D.\ Bekenstein: {\it Developments
    in General Relativity, Astrophysics and Quantum Theory:  A Jubilee
    in Honour of Nathan Rosen}, eds. F.I. Cooperstock, L.P. Horwitz,
    J. Rosen, p.\ 155, Bristol: IOP Publishing (1990)
\bibitem [48]{bs94} J.D.\ Bekenstein, 
    R.H.\ Sanders: \apj 429, 480 (1994)
\bibitem [49]{jdb92} J.D.\ Bekenstein:
    {\it Proceedings of the Sixth Marcel  Grossman Meeting on
    General Relativity}, eds. H. Sato \& T. Nakamura, p.\ 905,
    Singapore: World Scientific (1992)
\bibitem [50]{ni72} W.-T. Ni: \apj, 176, 769 (1972)
\bibitem [51]{rhs97} R.H.\ Sanders: \apj 480, 492 (1997)
\bibitem [52]{lln74} D.L.\ Lee, A.P.\ Lightman, W.-T. Ni: \pr, D10,
  1685 (1974)
\bibitem [53] {scdeal} C.\ Skordis, D.F.\ Mota, P.G.\ Ferreira, C.\ Boehm:
   astro-ph/0505519 (accepted PRL 2005)
\bibitem [54] {zheal} H.-S.\ Zhao, D.J.\ Bacon, A.N.\ Taylor, K\ Horne:
    \mnras (in press, 2006), astro-ph/0509590
\bibitem [55] {rhs89} R.H.\ Sanders: \mnras, 241, 135 (1989)
\bibitem [56]{zf05} H.-S. Zhao, B. Famaey:  \apj, 638, L9,
     astro-ph/0512435 (2005)
\bibitem [57]{will} C.M. Will: {\it Living Rev.\ Rel.}, 4,
    4 (2001)
\bibitem [58]{gian05} D.\ Giannios, \pr, D71, 103511 
   [astro-ph/0502122] (2005)
\bibitem [59] {mnv96} 
    J.\ M\'uller, K.\ Nordtvedt, D. Vokrouhlick\'y: \pr, D54,
    5927 (1996)
\bibitem [60] {ander} J.D.\ Anderson, et al.:
    {\it Phys.Rev.Lett.}, 81, 2858 (1998)
\bibitem [61] {rhs06} R.H. Sanders: \mnras, submitted (2006)
\bibitem [62] {bs05} J.D.\ Bekenstein, R.H.\ Sanders: astro-ph/0509519,
    (2005)
\bibitem [63] {prs} W.H.\ Press, B.S.\ Ryden, D.N.\ Spergel:
     1990, {\it Phys.Rev.Lett.}, 65, 1084 (1990)
\bibitem [64]{m94a} M.\ Milgrom: {\it AnnalsPhys} 229, 384 (1994)
\bibitem [65]{m99} M.\ Milgrom:  {\it Phys. Lett. A}
    253, 273 (1999)
\bibitem [66] {m05} M.\ Milgrom: astro-ph/0510117 (2005)
\end{thebibliography}
\end{document}